\documentclass{moriond}
\usepackage{natbib}

\def\be{\begin{equation}}
\def\ee{\end{equation}}
\def\bea{\begin{eqnarray}}
\def\eea{\end{eqnarray}}

\newcommand{\Photo}{}

\begin{document}
\vspace*{4cm}
\title{Einstein Telescope and Cosmic Explorer}

\author{Matteo Di Giovanni \\ on behalf of the ET collaboration}

\address{Dipartimento di Fisica, Università di Roma La Sapienza \\ and INFN Sezione di Roma 1 \\ Piazzale Aldo Moro 2\\
00185 Rome, Italy}

\maketitle\abstracts{
The goal of this talk is to give an overview of the current status of the development of the Einstein Telescope and Cosmic Explorer ground based gravitational wave (GW) detectors and of their foreseen scientific goals. These detectors will be up to a factor 8 more sensitive across the band covered by current detectors, namely LIGO, Virgo and KAGRA, and will extend the accessible frequency band towards the low frequency regime, i.e., below 10 Hz. These improvements will not only enhance the number and quality of GW observations, but will also enable researchers to have access to sources and physical processes which are out of reach for current detectors and explore the possibility of detecting previously unknown GW sources. The improvement in sensitivity in the low frequency regime will also increase the observation time of compact binary coalescence events, strengthening the collaboration with electromagnetic observatories for multimessenger observations of binary neutron star events. In fact, current detectors proved that joint observations of GW events with electromagnetic observatories are not only possible, but they can also give us unprecedented insights on the underlying physics of astrophysical processes.}

\section{Introduction}

\begin{figure}
\begin{minipage}{0.50\linewidth}
\centerline{\includegraphics[width=0.9\linewidth]{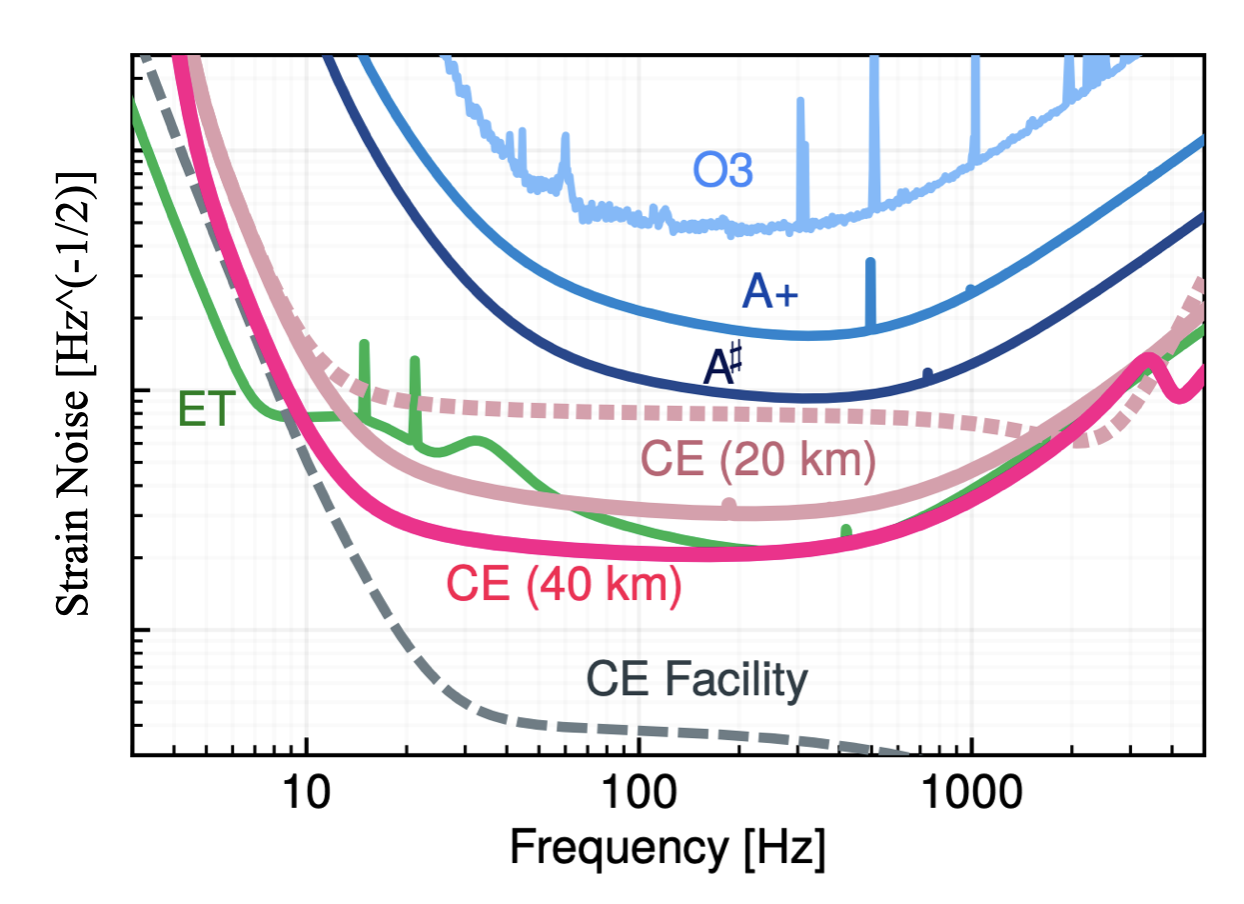}}
\end{minipage}
\hfill
\begin{minipage}{0.50\linewidth}
\centerline{\includegraphics[width=0.7\linewidth]{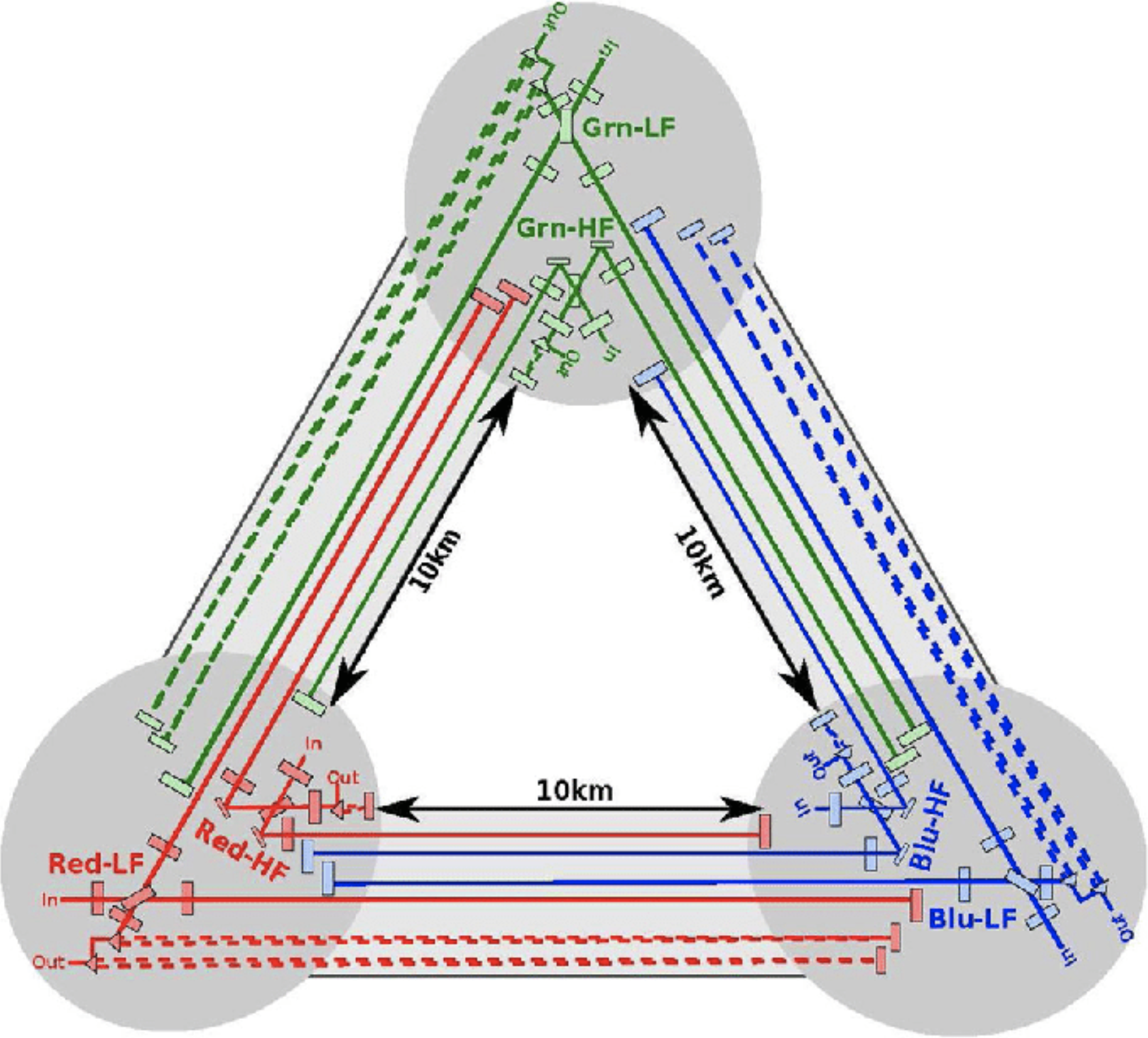}}
\end{minipage}
\caption[]{(left) ET and CE design sensitivity curves compared against design sensitivities of current and future GW detectors (LIGO A+ and A\# will be two implementations of the current Advanced LIGO detectors \cite{LIGOpost}, figure taken from \cite{CE4}). (right) Scheme of the ET triangular configuration \cite{ET2010}.}
\label{fig:comp}
\end{figure}

\begin{figure}
\begin{minipage}{0.50\linewidth}
\centerline{\includegraphics[width=0.7\linewidth]{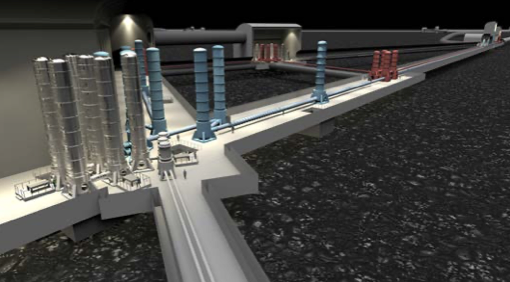}}
\end{minipage}
\hfill
\begin{minipage}{0.50\linewidth}
\centerline{\includegraphics[width=0.7\linewidth]{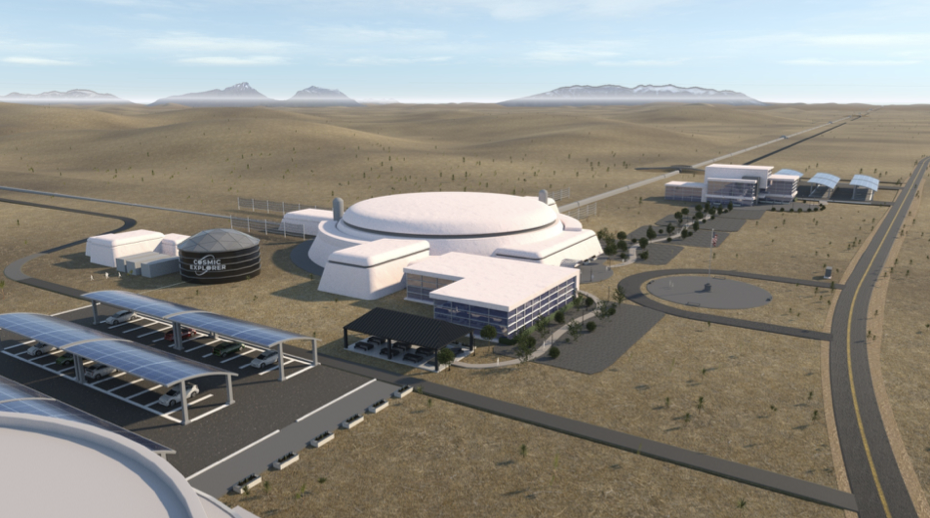}}
\end{minipage}
\caption[]{Artistic views of ET \cite{ET2020} and CE \cite{cesite}.}
\label{fig:art}
\end{figure}

Over the past few decades, gravitational wave (GW) astronomy has undergone a swift evolution, transitioning from resonant bar detectors, which were the GW detector of choice from the early 1970s up to the end of the 1990s, \footnote{Most of the resonant bar detectors were decommissioned only in the second decade of the XXI century, after supporting the observing runs of the early interferometers. The AURIGA detector is now on display in the premises of the Laboratori Nazionali di Legnaro in Italy.} to laser interferometers which became the preferred detector of the GW community since the beginning of the XXI century. The first generation of interferometers (LIGO, Virgo and GEO600), operational during the first decade of the century, proved that such detectors could reach their design target and could be effectively operated. The second-generation detectors, the currently operational Advanced detectors, namely Advanced LIGO \cite{aLIGO}, Advanced Virgo \cite{aVirgo} and KAGRA \cite{kagra}, have made groundbreaking discoveries in ten years of observations, with over 90 events detected from the observing runs O1 to O3 \cite{gwtc1, gwtc2, gwtc3} and 200 more public alerts in O4. The current generation of detectors has validated key astrophysical phenomena, like the fact that compact objects mergers within one Hubble time exist and can be observed by ground based detectors. They also paved the way for multimessenger astronomy, as evidenced by GW170817 \cite{Branchesi2018} and the successful observational campaigns with partner observatories that followed. However, the full scientific potential of GW observations remains untapped, especially in the low-frequency domain, post-merger dynamics, detection of continuous GW from isolated pulsars and the stochastic GW background of both astrophysical and cosmological origin. Next generation observatories, namely the Einstein Telescope (ET) \cite{ET2010, ET2011, ET2020} and Cosmic Explorer (CE) \cite{CE1, CE2, CE3, CE4} (Figures \ref{fig:comp} and \ref{fig:art}), promise to revolutionize, in the next decade, the field by vastly enhancing sensitivity and the accessible bandwidth, extending our view of the universe to unprecedented depths and details.

\section{The vision of next generation GW detectors}
\subsection{Einstein Telescope}
Einstein Telescope is the European proposal for a next generation GW detector, first proposed in 2010 \cite{ET2010}. The foreseen improvements with respect to current-generation detectors include the extension of the observation bandwidth from the current limit of about 20\,Hz to 2\,Hz and an improvement of the sensitivity up to a factor 8 across the band covered by current detectors \citep{ET2020} (Figure \ref{fig:comp}). To reduce seismic motion at the input of the suspension system of the mirrors and to reduce the impact of atmospheric disturbances \citep{hutt} and Newtonian noise (NN), ET is also foreseen to be built underground at a currently planned depth between 250 m and 300 m. For what concerns the detector configuration, there are two proposals currently under consideration. The most recent one is that of a detector network composed by two widely separated L-shaped detectors with 15 km long arms \cite{coba, Iacovelli_2024}. On the other hand, the original project foresees three pairs of nested interferometers arranged in an equilateral triangle (Figure \ref{fig:comp}) with the sides 10\,km long \cite{et, ET2010, ET2011, ET2020}. For each interferometer pair, one detector is optimized for low frequencies (2\,Hz $< f <$ 40\,Hz) and the other for high frequencies ($f>$ 40\,Hz, also called xylophone configuration). Since recently, moving the lower limit to 3 Hz is being considered as well \citep{ET2020}. In both the 2L and the triangle configurations, ET is expected to be hosted underground. References \cite{coba, Iacovelli_2024} also found that the difference between the two configurations in terms of the reachable science goals is minimal.

Generally speaking, the extension of the bandwidth to 2\,Hz and the sharp increase in sensitivity will significantly improve the rate of detected events giving the possibility to issue early warnings for the coalescence of compact objects (CBC) several minutes, if not hours depending on the source, before the merger \cite{Branchesi_2016,Maggiore_2020,Nitz_2021,coba, Hu_2023}. In fact, with respect to current detectors, compact binary coalescence (CBC) signals will spend more time in the ET and CE accessible bandwidth, therefore enabling early detection and accurate sky localization.

\subsection{Cosmic Explorer}
CE was proposed by American institutions a few years after ET and, since its first proposal, is expected to follow a more classical approach. It foresees the construction of a network of two surface built widely separated L-shaped interferometers, one with 40 km long arms and the other with 20 km long arms \cite{CE2,CE3,nsfreport}, although it considers alternate scenarios in which CE consists of a single 40 km facility, two 40 km facilities, or two 20 km facilities \cite{CE3}. The baseline configuration will be widely based on LIGO technology, including 1064 nm wavelength laser and, contrary to ET, room temperature operations. Other options are also under consideration for future upgrades. Being of surface, it is also expected to adopt appropriate noise suppression and isolation systems to reach the design sensitivity goals at low frequency.

The call for a 20 km facility, in addition to a 40 km facility, was motivated by the necessity of measuring the post-merger oscillations of binary neutron star (BNS) mergers which are expected to happen between 2–4 kHz, where the optical response of a 40 km detector is reduced due to the travel time of the light down the arms \cite{CE3}. In particular, the 20 km could make use of “tuned” operation, i.e., the quantum noise is reduced in the 2–4 kHz band at the expense of higher quantum noise elsewhere. It was also found that 20 km detector yielded a 30$\%$ improvement in average signal-to-noise ratio in the 2–4 kHz band compared to a 40 km detector, which would better address most science themes anyway for its superior noise performance outside the 2–4 kHz frequency range \cite{CE3}.

\subsection{ET and CE global network}

Although not matching the current ET design sensitivity at very low frequencies (Figure \ref{fig:comp}), CE complements ET in terms of global coverage and redundancy. The collaborative operation of ET and CE as a global 3G network will substantially improve event localization, potentially with an uncertainty of 1 square degree for BBH and 10 square degress for BNS \cite{Iacovelli_2024,bidbook}, critical for triggering electromagnetic follow-ups in multimessenger astronomy. CE will also contribute significantly to key science goals: constraining merger rates over cosmic time, probing formation channels of supermassive and intermediate-mass black holes, and detecting potential primordial black hole populations.

\section{Challenges for next generation GW detectors}
As already mentioned, future 3rd generation GW detectors will be much more sensitive at frequencies below 20 Hz, with respect to current detectors. Among the other things, this increase in sensitivity will  be beneficial for the observations of intermediate mass black holes (IMBH) and to trigger multimessenger observation campaigns for BNS mergers with great advance. In particular, BNS signals can spend up to 20 hours below at low frequency making this frequency band crucial for early warnings, accurate sky localization and to effectively exploit multi-messegner observational campaigns. 

Therefore, any degradation with respect to design in the low-frequency sensitivity of ET may significantly hinder the capability of early detections and multimessenger observations for BNS mergers and may reduce the quality of the observations for IMBH \cite{Nitz_2021, coba, digiovanni2025}. As a consequence, since seismic disturbances, of both natural (see, e.g., \cite{acernese2004properties, virgo2006, o3noise}) and anthropogenic origin (e.g., \cite{acernese2004properties, virgo2006, saccorotti, piccinini, o3noise}), are the main source of noise limiting the detector sensitivity at low frequency and can affect GW data in many ways (e.g., \citep{Fiori2020, o3noise, saulson, harms}), seismic characterization studies at the candidate sites to host ET are paramount \cite{digiovannietal2021, digiovanni2023, digiovanni2025}. The goal is to guarantee a suitable environment for this future detector that makes the reaching of its design sensitivity possible \citep{amann} through appropriate design of noise suppression systems. 

Among all the possible noise sources, the trickiest to tackle is Newtonian Noise (NN) \cite{saulson, harms}. NN originates from ground motion and atmospheric disturbances that alter the dynamic mass density distribution around the test masses and the gravitational field felt by the masses themselves (flowing water masses may also influence NN). This means that NN is a gravitational interaction and can be neither physically nor mechanically shielded. Appropriate filters for noise subtraction and design of environmental sensors arrays and low ground motion environments are needed. Since seismic ground motion attenuates with depth, NN is expected to be less prominent underground. For this reason, ET will be built underground whereas CE, which will be built on surface, will adopt appropriate noise suppression techniques at the cost of being slightly less sensitive than ET at the lower end of the accessible frequency band (Figure \ref{fig:comp}).

As far as NN suppression is concerned, the most complete study issuing robust and realistic forecasts about the achievable NN mitigation factors in ET can be found in Ref. \cite{badaracco2019}. In particular, it reports that the cancellation of NN from a body wave seismic field can be achieved by a factor between 2 and 3. This result was achieved after simulating the noise cancellation process with 15 seismometers per test mass in a plane and assuming an isotropic and body wave field. Ref. \cite{badaracco2019} concluded that a reduction of up to a factor 10 could be possible as well, but only at a precise frequency, i.e., the frequency for which we choose to optimize the NN cancellation sensor array. For example, if the sensor array is optimised for NN cancellation at 15 Hz, the NN reduction at other frequencies is only a factor of 2. Moreover, if the NN cancellation is performed broadband, the reduction factor will never exceed 2 or 3 depending on the frequency. Since many astrophysical events span a wide range of frequencies, we can conclude that a realistic forecast for NN cancellation in next generation detectors will hardly exceed a factor 3.
Using an adaptive Wiener filter for NN cancellation, Ref. \cite{Koley2024} also concluded that a factor 10 could be possible, but only for noise field fluctuations over long time scales. For minute-long fluctuations, the reduction factor is only 2.4. 

In general, NN reduction is expected to be a huge computational and technological effort and the suppression factor will strongly depend on the infrastructure. For example, reaching a factor 3 suppression factor for NN in ET would require a few tens of seismometers placed in boreholes around each test mass in 3D \cite{badaracco2019}. Since ET will have 12 test masses, we are talking about 120 sensors for which 3D data have to be processed, let alone the fact that we assume that the boreholes where the sensors are placed exactly where expected.




\begin{figure}
    \centering
    \includegraphics[width=0.5\linewidth]{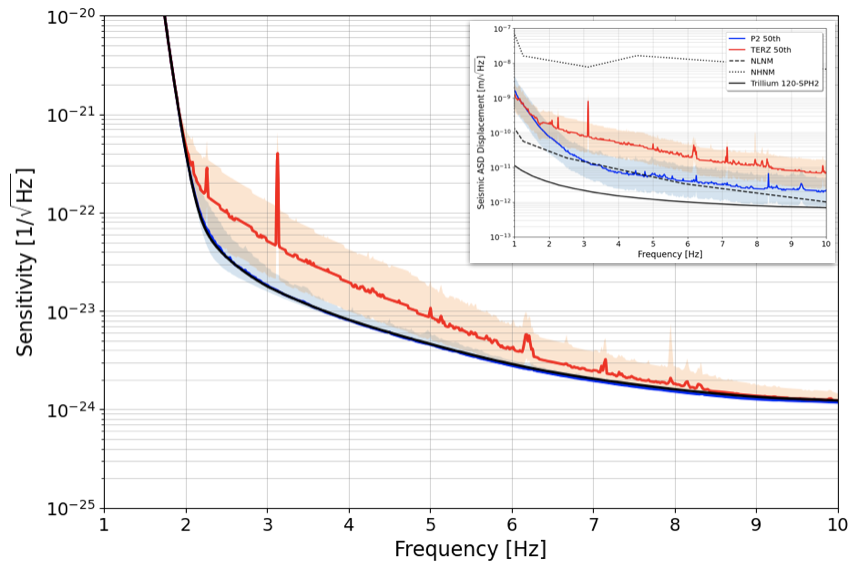}
    \caption{Effects of the site dependent ambient noise over the sensitivity of ET from \cite{digiovanni2025}. The black curve represents the design sensitivity from \cite{coba}; the blue line is the resulting sensitivity at the Sardinia candidate site; the red curve is the resulting sensitivity at the currently available Terziet site. The plot insert shows the difference in seismic noise levels recorded at the two sites.}
    \label{fig:sens}
\end{figure}
\section{The importance of site quality}
At the moment, the CE collaboration has not yet identified potential candidate sites for its detectors, but is developing the criteria for selecting them \cite{cesite}. On the other hand, two primary candidates are under evaluation for the location of ET: the Euregio Meuse-Rhine (EMR) area, at the border between Belgium and the Netherlands and represented by the village of Terziet, and the Sos Enattos area in Sardinia, Italy. Both sites have been the subject of thorough characterization studies \cite{naticchionietal2014, naticchionietal2020, digiovannietal2021, alloccaetal2021, digiovanni2023,digiovanni2025, saccorotti23,koley19, Bader_2022, koley2022surface} with goal of assessing their suitability to host ET. A third candidate in Lausitz, Germany, is also entering the site selection process. 

Site quality, particularly in terms of seismic activity and low NN environments, will play a critical role in the operations of ET. And, recently, some studies started to investigate the impact of site dependent noise over the performance of ET \cite{alloccaetal2021,janssens2024,digiovanni2025}. Notably, \cite{digiovanni2025} shows that Sardinia exhibits noise characteristics closer to the ET requirements even without noise suppression factors, whereas an hypothetical detetector located at the currently available site of Terziet will require robust and complex NN mitigation systems, since the effects of local ambient noise over the design sensitivity of ET are apparent (Figure \ref{fig:sens}).  Using an approach which joins ambient noise studies and simulations of astrophysical events, \cite{digiovanni2025} also shows that the difference of the effects on the ET design sensitivity curve reflects on the performance of ET for early warning purposes in the low frequency band as well.

\section{Conclusions}

The development of the Einstein Telescope and Cosmic Explorer represents a pivotal step forward in the evolution of GW astronomy. These next generation observatories will greatly expand the accessible frequency range and significantly increase sensitivity, enabling the detection of sources and physical processes beyond the reach of current detectors. While the novel configurations and ambitious technological improvements pose considerable scientific and engineering challenges, addressing these issues is essential to fully realize the potential of the new detectors. By overcoming environmental and infrastructural obstacles, the community will unlock new opportunities for early detection, precise localization, and multimessenger follow-up of astrophysical events. The synergistic operation of ET and CE as a global network will mark the beginning of a new era in gravitational wave science, offering deep novel insights into the dynamics of the Universe.

\bibliography{moriond}


\end{document}